\documentstyle[12pt,a4]{article}
\begin{document}
\begin{titlepage}
\begin{flushright}
hep-th/9507034 \\
July 1995
\end{flushright}
\vspace*{\fill}
\centerline{\huge\bf Anyon trajectories}
\centerline{\ }
\centerline{\huge\bf and the systematics}
\centerline{\ }
\centerline{\huge\bf of the three-anyon spectrum}

\vspace{2cm}
\centerline{\large Stefan Mashkevich\footnote{~E-mail:
mash@phys.unit.no}}
\vspace{.2cm}
\centerline{\large Institute for Theoretical Physics}
\vspace{.1cm}
\centerline{\large 252143 Kiev, Ukraine}

\vspace{0.75cm}

\centerline{\large
Jan Myrheim\footnote{~E-mail: janm@phys.unit.no},
K{\aa}re Olaussen\footnote{~E-mail: kolausen@phys.unit.no},
Ronald Rietman\footnote{~E-mail: rietman@prl.philips.nl.
Permanent address: Philips Research Laboratories,
NL-5656 AA Eindhoven, The Netherlands}
}
\vspace{.2cm}
\centerline{\large Institutt for fysikk, NTH,}
\vspace{.1cm}
\centerline{\large Universitetet i Trondheim}
\vspace{.1cm}
\centerline{\large N--7034 Trondheim, Norway}

\vspace*{\fill}
\vspace{.5cm}
\begin{abstract}
We develop the concept of trajectories in anyon spectra,
i.e., the continuous dependence
of energy levels on the kinetic angular momentum.
It provides a more economical and unified description,
since each trajectory contains an infinite number of
points corresponding to the same statistics.
For a system of non-interacting anyons in a harmonic potential,
each trajectory consists of two infinite straight line segments,
in general connected by a nonlinear piece.
We give the systematics of the three-anyon trajectories.
The trajectories in general cross each other at the bosonic/fermionic
points.
We use the (semi-empirical) rule
that all such crossings are true crossings, i.e.\ the order
of the trajectories with respect to energy
is opposite to the left and to the right of a crossing.

\end{abstract}

\vspace*{\fill}

\end{titlepage}

\section{The concept of trajectories}

Anyons \cite{Leinaas77,Wilczek82-1} are two-dimensional
particles whose wave function $\psi$ obeys the interchange conditions
\begin{equation}
P_{mn} \psi = \exp(i\pi\nu) \psi,
\label{int}
\end{equation}
where $P_{mn}$ denotes continuous anticlockwise interchange
of particles $m$ and $n$, such that no other particles are
encircled, and
$\nu$ is the statistics parameter, which may be any real number.
For a given Hamiltonian, not explicitly dependent on $\nu$
and with a discrete spectrum for any $\nu$,
the energy eigenvalues $E_k(\nu)$ and (usually) the corresponding
eigenfunctions $\psi_k(\nu)$ will be continuous functions of $\nu$.
A continuous function $E_k(\nu)$, for a fixed index $k$ and
$\nu \in \langle -\infty,\infty \rangle$, is what we will
refer to as an (anyon) trajectory.
For an earlier discussion of the $\nu$ dependence
of the states, see \cite{Karl}.

Since the exponent in (\ref{int}) is the only place where $\nu$
appears in this formulation,
the spectrum and the set of energy eigenstates
are obviously periodic in $\nu$ with period 2; thus, all the
information is contained in the sets $\{E_k(\nu)\}$
and $\{\psi_k(\nu)\}$ for $\nu \in [0,2\rangle$.
However, individual trajectories are not generally periodic in $\nu$;
one way to see this is to observe that when $\nu$ increases
continuously, the kinetic angular momentum $L$ changes
according to the formula
\begin{equation}
L(\nu) = L(0) + \frac{1}{2} N(N-1)\nu,
\label{angm}
\end{equation}
with $L(0)$ an integer\footnote{
This formula can be proved by noting that a $2\pi$ rotation of
the whole system multiplies the wave function by a phase factor,
which equals, on the one hand, $\exp[2i\pi L(\nu)]$, and on the other
hand, $\exp[N(N-1)i\pi\nu]$,
since each of the $N(N-1)/2$ pairs of anyons is
interchanged twice.},
so that changing $\nu$ by 2 will not bring us back to the same state.
It is assumed that the Hamiltonian is rotationally invariant, so
that $L$ is a good quantum number.
Thus, $E_k(\nu+2\ell)$, where $\ell = 0,\pm 1,\pm 2,\ldots$,
will be the energy of some state with statistics
parameter $\nu$, but the index of that state will be
different from $k$:
\begin{equation}
E_k(\nu+2\ell) = E_{k'}(\nu);
\label{period}
\end{equation}
where $k'$ depends on $k$ and $\ell$: $k'=k'(k,\ell)$.
At any given $\nu$, the trajectories $k$ and $k'$ look different,
having different angular momenta $L_k(\nu)$ and
$L_{k'}(\nu) = L_k(\nu+2\ell) = L_k(\nu) + N(N-1)\ell$,
but when viewed on the interval
$\langle-\infty,\infty\rangle$,
they are seen to be copies of one and the same
trajectory, shifted by $2\ell$ along the $\nu$ axis.

To represent the trajectories in a way which is not redundant,
it is convenient to take $L$ as a parameter instead of $\nu$.
$L$ is directly related to $\nu$ by equation (\ref{angm}),
it is an observable (gauge invariant) quantity, and
the dependence $E(L)$ is the same for all the trajectories
that are shifted copies of one another.
We arrive at the viewpoint that the states of
an $N$-anyon problem may be obtained from the set
of trajectories, i.e., functions $E(L)$ for
$L\in\langle-\infty,\infty\rangle$.
This provides a more economical and unified way of decribing the
$N$-anyon spectra.
For each trajectory there is an integer value $L(0)$ such that
the statistics is bosonic at angular momenta $L=L(0) + N(N-1)\ell$.
We may always choose $-N(N-1)/2 < L(0) \leq N(N-1)/2$. The value of
$L(0)$ groups the
set of trajectories into $N(N-1)$ classes. Only trajectories
from the same class can cross, i.e.\ have the same $E$ and $L$ at
the same statistics.

A part of a trajectory
between any two neighboring bosonic points corresponds
to what is usually referred to as one anyon state
on the interval $\nu \in [0,2\rangle$.
It is useful to have a geometric picture of this:
If one plots a trajectory for $-\infty < L < \infty$
and then wraps the plot around a cylinder of circumference
$N(N-1)$, then points which correspond to $\nu$ differing by an even
number (that is, to the same statistics) will
fall on the same vertical line.
On the surface of the cylinder one will see the set
of all $N$-anyon states corresponding to the trajectory.
Thus, an infinite number of pieces of trajectories
(or states in the usual terminology)
on the interval $\nu \in [0,2\rangle$
are shown to make up one single trajectory on the interval
$\nu \in \langle -\infty,\infty\rangle$.
In other words, certain anyon
(and in particular, boson or fermion) states
that are usually considered entirely different,
are in fact parts of one and the same continuous
pattern. This is reminiscent of the method of organizing
the spectra of particle physics into Regge trajectories.
In fact, by applying the concept of Regge trajectories to
a two-dimensional, two-particle system, one obtains the same grouping
of bosonic/fermionic states.

It follows that if some trajectory $E(L)$ is analytic, then
it is in principle sufficient to find only one
of its pieces. The rest is uniquely determined by
analytic continuation. This means that
it is sufficient to find one anyonic state in some range of
$\nu$ and then an infinite number of others are obtained
``automatically'' by analytic continuation and periodicity.
However, trajectories are not always analytic.
With our assumption that the Hamiltonian does not explicitly
depend on $\nu$ (and involves only non-singular
interactions) both energies and wave functions will
depend analytically on $\nu$, except possibly at bosonic points.
The source of non-analyticity is
that the relative angular momentum of one pair of
particles becomes zero, which can only happen at bosonic
points. When there is non-analytic behavior at some point
with degeneracy, one must determine how each trajectory
continues through this point by investigating the corresponding
wave functions. They should change continuously across these points.

\section{Two- and three-anyon trajectories}

We are going now to apply this reasoning to the problems of
two and three anyons in a rotation symmetric
harmonic potential.
We scale the variables so that $\hbar = 1$, the mass $m=1$
and the angular frequency $\omega=1$.
Two more preliminary remarks are in order.
First, the potential is parity invariant; hence,
if $\psi$ is an eigenfunction of the Hamiltonian
and satisfies (\ref{int}), then its complex conjugate
$\bar{\psi}$ is an eigenfunction of the Hamiltonian
with the same energy and satisfies (\ref{int}) with $-\nu$
instead of $\nu$. In particular, this implies that all
information about the states is in fact contained
in the interval $\nu\in[0,1]$.

Second, there is the tower structure of the spectrum
\cite{Sen,SporreNP}.
It has been observed that, for any number of
anyons, all the states come in towers, the angular momentum
being the same for all members of a tower and the energies
being $E(\nu)$, $E(\nu)+2$, $E(\nu)+4, \ldots$,
where $E(\nu)$ is the energy of the lowest, ``bottom'' state.
This sequence of levels is due to radial excitations.
The radial coordinate $r$ is defined by
$r^2 = \sum_i\, (x_i-X)^2+(y_i-Y)^2$,
with $(X,Y)$ the center-of-mass coordinates.
Thus, it is sufficient to find only the bottom states
(consequently, bottom trajectories); from now on, we will
always mean these, unless otherwise specified.

Now we will demonstrate that in the two-anyon problem there
is only {\em one\/} (bottom) trajectory,
namely the continuation of the ground state.
Recall that the one-particle spectrum
in the harmonic potential consists of states with
$E=1,2,3,\ldots$ and $L=E-1, E-3, \ldots, -(E-3),-(E-1)$
(so the degeneracy equals the energy).
Define complex coordinates by $z_j = (x_j+i y_j)/\sqrt{2}$.
Since the center-of-mass motion is trivial, we will always
concentrate on the relative motion only. For two anyons, the
complete set of (not normalized) solutions may be written as
\begin{equation}
\psi_{ln}(z,\bar{z}) = \tilde{z}^{|2l+\nu|}\,
{}_1\!F_1 \left(-n,|2l+\nu|+1;z\bar{z}\right)
\exp\left(-z\bar{z}/2\right)
\label{psiln}
\end{equation}
with $z=z_1-z_2$, $l$ an integer,
$n$ a non-negative integer, and $\tilde{z}$ standing for $z$
if $2l +\nu \geq 0$ and for $\bar{z}$ if $2l +\nu<0$;
the energy and the angular momentum are
\begin{equation}
E_{ln}(\nu) = |2l+\nu| + 2n + 1, \qquad
L_{ln}(\nu) =  L_l(\nu) = 2l+\nu,
\end{equation}
respectively. Towers consist of states with the same $l$ and
different $n$; in accordance with the aforesaid, it is sufficient
to consider the bottom states only, for which $n=0$ and
\begin{equation}
E = |L|+1
\label{EL}
\end{equation}
{\it irrespective\/} of $l$. Indeed, all the states
in question belong to one and the same trajectory,
because the relation (\ref{period}) does hold,
in the form $E_0(\nu+2l) = E_l(\nu)$ (and the same for the
wave function).
For $-1<\nu<1$ this is the ground state,
\begin{eqnarray}
\psi_0(z,\bar{z}) &\!\!\! = \!\!\!& \tilde{z}^{|\nu|} \exp
\left(-z\bar{z}/2\right), \label{psigr} \\
E_0(\nu) &\!\!\! = \!\!\!& |\nu| + 1. \label{Egr}
\end{eqnarray}
Thus, {\it all\/} bottom states are continuations
of {\it one\/} ground state.
Equation (\ref{EL}) exemplify our previous remark
that a trajectory may become non-analytic when the
relative angular momentum of one pair of particles
(here $L$ itself) becomes zero. By adding the
pair potential $g^2/\vert z \vert^2$ to the Hamiltonian,
(\ref{EL}) is changed to $E(L)=\sqrt{L^2+g^2}+1$,
which demonstrates that singular interactions may lead to
different behaviours.

We go now to the problem of three anyons, which is extremely
interesting due to its nontriviality, on the one hand, and
the possibility of a more or less exact analysis, on the
other hand. Recall some results available.
A state starting with $(E,L)$ at $\nu=n$ ($n$~integer)
may reach one of the following points at $\nu=n+1$:
$(E+3,L+3)$, $(E+1,L+3)$, $(E-1,L+3)$, $(E-3,L+3)$
\cite{Sporre,Illum}.
In other words, in any interval from a bosonic to a fermionic
point or from a fermionic to a bosonic point, each state is
characterized by an (average) slope
\begin{equation}
s = \frac{\Delta E}{\Delta \nu} = 3\frac{\Delta E}{\Delta L}
= \; \pm 1 \;\;\; \mbox{or} \;\;\,\pm 3,
\end{equation}
where $\Delta E$ is the change in energy, and $\Delta L = 3\Delta \nu
=3$
is the change in angular momentum. For the $s = \pm 3$ states
the dependence $E(L)$ is linear, so that $s=dE/d\nu$ is the slope
at any point in the interval, and their wave functions may be written
down exactly \cite{Dunne,Chou,Mash92}. For the $s = \pm 1$ states
the dependence is nonlinear, so a slope of $\pm 1$ is indeed only an
average slope. Now, there are
exact expressions for the multiplicities of states with all
slopes \cite{Mash93}. If $\tilde{b}^n(E,L)$
denotes the number of bottom states in the relative motion
spectrum that go from $(E,L)$ to $(E+n,L+3)$ as
$\nu$ goes from $2m$ to $2m+1$, i.e., from a bosonic to a
fermionic point, then there are the asymptotic
expressions for $E,L \gg 1$ (terms of order unity being omitted)
\begin{eqnarray}
\tilde{b}^{+3}(E,L) &\!\!\!=\!\!\!& \frac{3L - E}{12}
 \phantom{\frac{2E - \left|3L-E\right|}{12}}
 \left(\phantom{-}\frac{E}{3} < L < E\right), \label{b+3} \\
 \nonumber \\
\tilde{b}^{+1}(E,L) &\!\!\!=\!\!\!&
 \frac{2E - \left|3L-E\right|}{12}
 \phantom{\frac{3L - E}{12}}
 \left(-\frac{E}{3} < L < E\right), \label{b+1}
\end{eqnarray}
and the exact formulas
$\tilde{b}^{-1}(E,L) = \tilde{b}^{+1}(E-1,-L-5)$,
$\tilde{b}^{-3}(E,L) = \tilde{b}^{+3}(E-6,-L-6)$.
In each of the formulas, $L$ has to be within the
respective interval specified and the equality
$ L \equiv E$~(mod 2) has to hold, otherwise
$\tilde{b}^n(E,L)=0$. Recall also that for
all states at any statistics, there is the inequality
$|L| < E$ \cite{Chitra}.

As one goes along a trajectory,
the slope will change at certain values of $L$,
so any trajectory possesses a sequence of slopes
and points of slope change.
We make now the following statements.
\begin{itemize}
\item[(i)] For each trajectory there exists an $L_+$ such that
$s=+3$ if and only if $L>L_+$.
\item[(ii)] For each trajectory there exists an $L_-$ such that
$s=-3$ if and only if $L<L_-$.
\item[(iii)] The change of slope from/to $\pm3$ can occur at
bosonic points only, while the change from $-1$ to $+1$ or vice
versa can occur at both bosonic and fermionic points.
\end{itemize}
To prove statement (i), note first that the inequality
$E > L$ and the fact that $dL/d\nu = 3$ force every trajectory
to have $s=+3$ at least for
some large positive values of $L$.
Now, the explicit form of the $s=+3$ states
\cite{Wu84,Dunne,Chou} is such that if one of them exists for
some $\tilde{\nu}$, it does exist
with $s=+3$ for any $\nu > \tilde{\nu}$,
that is for any $L > \tilde{L} = L(\tilde\nu)$;
in other words, if $s=+3$ at some point, then
$s=+3$ everywhere to the right of that point. This
concludes the proof.
A ``mirror reversed'' reasoning proves statement (ii).

Statement (iii) is proved upon noticing that the change
of slope from/to $\pm3$ always implies nonanalyticity in
the function $E(L)$, because it is a change between a
linear and a nonlinear dependence. Such nonanalyticity
can only happen at bosonic points, as discussed in Sec.\ 1.
This is also associated with the breakdown
of regular perturbation theory\footnote{
The usual way to introduce perturbation theory is by
replacing the anyonic boundary
conditions by a $\nu$ dependent Aharonov-Bohm-type
(perturbative) interaction.}
at such points \cite{Chou,pert}.

Note that since statements (i) and (ii) mean that each trajectory
has $s=-3$ for some values of $L$ and $s=+3$, for some
others, it follows that each trajectory
is indeed nonanalytic at least at one point.
(This is of course equally true for the $N$-anyon problem,
where the extreme slopes, with linear behavior, are
$\pm N(N-1)/2$, and each trajectory possesses both of
them just like in the case at hand; cf.~\cite{Karl}.)

So far we may conclude that the behavior of each trajectory
has the following features: At large negative $L$,
slope $-3$, at some point $(E_-,L_-)$ a change to $\pm1$,
for $L_- < L < L_+$ some sequence of slopes $\pm1$ and
finally a change to $+3$ at $(E_+,L_+)$ and always $+3$
at $L > L_+$. Note that there is exactly one trajectory
for which $L_-=L_+$, that is which has slopes
$+3$ and $-3$ only: the one which contains the
ground state near Bose statistics,
\begin{equation}
   \psi_0  = \left\{ \begin{array}{ll}
   (\bar{z}_{12} \bar{z}_{23} \bar{z}_{31})^{-\nu} \qquad & \mbox{for
}
   \nu < 0, \\ \\ (z_{12}z_{23}z_{31})^{\nu} \qquad &
   \mbox{for } \nu \geq 0, \end{array} \right.
\end{equation}
where $z_{jk} = z_j - z_k$, and the overall Gaussian
factor is understood.

We will now prove that
\begin{itemize}
\item[(iv)] Each bottom trajectory may be unambiguously labeled by
its $(E_+,L_+)$ point.
\end{itemize}
Indeed, given an arbitrary bosonic point $(E,L)$, it is easy
to see that the number of bottom trajectories for which
$(E_+,L_+)=(E,L)$, equals
$$\tilde{b}^{+3}(E,L) - \tilde{b}^{+3}(E-6,L-6) \le 1,$$
where the inequality follows from eq.~(\ref{b+3}).
The equality is reached, meaning that a trajectory
with $(E_+,L_+)=(E,L)$ exists, for each $(E,L)$
such that $\frac{E}{3} < L < E$ and $L \equiv E$~(mod 2).
Like the equation itself, this
statement is true up to terms of the order of
unity. Fig.~1 shows the exact picture of the
distribution of the $(E_+,L_+)$ points on the plane
(the bullets and triangles will be explained
later).

\section{The linear parts of the trajectories}

We proceed to show how all the linear parts of the
trajectories (and thus all the three-anyon linear
states) are constructed as excitations of the ``ground
trajectory'', the one which contains the ground
state. By virtue of symmetry, it is enough
to consider $s=+3$ states only. It is convenient
to use the coordinates
proposed in \cite{3ansol}, which are the discrete Fourier
transform of the complex particle coordinates. The relative
motion is described by
\begin{equation}
  u=\frac{1}{\sqrt{3}}\left(
  z_1+\eta z_2+\eta^2 z_3\right),\qquad
  v=\frac{1}{\sqrt{3}}\left(
  z_1+\eta^2 z_2+\eta z_3\right),
\end{equation}
where $\eta=\exp\left({2i\pi}/{3}\right)=
(-1+i\sqrt{3})/{2}$.
Introduce the creation and annihilation operators
\begin{eqnarray}
a_u = \frac{1}{\sqrt{2}} (\bar{u} + \partial_u), & \qquad &
a_u^{\dagger} = \frac{1}{\sqrt{2}} (u - {\partial}_{\bar u}), \\
b_u = \frac{1}{\sqrt{2}} (u + {\partial}_{\bar u}), & \qquad &
b_u^{\dagger} = \frac{1}{\sqrt{2}} (\bar{u} - \partial_u),
\end{eqnarray}
and, with $u\to v$, the same relations for $a_v$, $a_v^{\dagger}$,
$b_v$, $b_v^{\dagger}$.
Then the relative Hamiltonian and angular momentum operator become
\begin{eqnarray}
H &\!\!\!=\!\!\!& a_u^{\dagger}a_u + a_v^{\dagger}a_v +
b_u^{\dagger}b_u + b_v^{\dagger}b_v + 2, \\
L &\!\!\!=\!\!\!& a_u^{\dagger}a_u + a_v^{\dagger}a_v -
b_u^{\dagger}b_u - b_v^{\dagger}b_v.
\end{eqnarray}
The commutation relations are $[a_u,a_u^{\dagger}]=
[b_u,b_u^{\dagger}]=[a_v,a_v^{\dagger}]=[b_v,b_v^{\dagger}]=1$,
and all other commutators vanish.
Consequently, if
$\psi$ is a common eigenstate of $H$ and $L$ with quantum
numbers $(E,L)$,
then $(a_u^{\dagger})^k (a_v^{\dagger})^l
(b_u^{\dagger})^m (b_v^{\dagger})^n\psi$
is also a common eigenstate with
$(E+k+l+m+n,L+k+l-m-n)$.
However, to yield true anyonic eigenstates, a combination of
the creation operators must be fully symmetric and
produce wave functions that are not singular at the points
where the positions of two particles coincide.

The pairs $(u,v)$ and $(\bar{v},\bar{u})$ define two equivalent
irreducible representations of the permutation group $S_3$:
for example, the pair $(u,v)$ transforms, under the
six possible permutations, into
$(u,v)$, $(\eta u,\eta^2 v)$, $(\eta^2 u, \eta v)$,
$(v,u)$, $(\eta v,\eta^2 u)$, $(\eta^2 v, \eta u)$, respectively.
Therefore fully symmetric quantities are those of the form
$u^k\bar{u}^lv^m\bar{v}^n + v^k\bar{v}^lu^m\bar{u}^n$
with $k-l-m+n \equiv 0\! \pmod{3}$.
The pairs $(a_u^{\dagger},a_v^{\dagger})$,
$(b_v^{\dagger},b_u^{\dagger})$, and $(\bar{v},\bar{u})$
all transform like $(u,v)$.
Taking successive products of the above-mentioned representations
of $S_3$ and decomposing them into irreducible representations,
it is straightforward to prove that all symmetric polynomials in
$a_u^{\dagger},a_v^{\dagger},b_u^{\dagger},b_v^{\dagger}$
can be expressed as polynomials in the following basic symmetric
polynomials:
\begin{equation}
\begin{array}{lclcl}
(k \; l \: m \, n) \\
(0 \; 3 \; 0 \; 0) : & \qquad &
c^{\dagger}_{3,-3} & = &
(b_u^{\dagger})^3+(b_v^{\dagger})^3, \\
(0 \; 1 \; 0 \; 1) : & \qquad &
c^{\dagger}_{2,-2} & = &
b_u^{\dagger}b_v^{\dagger}, \\
(0 \; 2 \; 1 \; 0) : & \qquad &
c^{\dagger}_{3,-1} & = &
(b_u^{\dagger})^2a_v^{\dagger}+(b_v^{\dagger})^2a_u^{\dagger}, \\
(1 \; 1 \; 0 \; 0) : & \qquad &
c^{\dagger}_{20} & = &
a_u^{\dagger}b_u^{\dagger} + a_v^{\dagger}b_v^{\dagger}, \\
(2 \; 0 \; 0 \; 1) : & \qquad &
c^{\dagger}_{31} & = &
(a_u^{\dagger})^2b_v^{\dagger}+(a_v^{\dagger})^2b_u^{\dagger}, \\
(1 \; 0 \; 1 \; 0) : & \qquad &
c^{\dagger}_{22} & = &
a_u^{\dagger}a_v^{\dagger}, \\
(3 \; 0 \; 0 \; 0) : & \qquad &
c^{\dagger}_{33} & = &
(a_u^{\dagger})^3+(a_v^{\dagger})^3.
\end{array}
\label{q}
\end{equation}
The meaning of the subscripts is that
\begin{equation}
[H,c^{\dagger}_{pq}]=p\,c^{\dagger}_{pq}, \qquad
[L,c^{\dagger}_{pq}]=q\,c^{\dagger}_{pq},
\end{equation}
so $c^{\dagger}_{pq}$ changes the energy by $p$ and the angular
momentum
by $q$. It remains to check for possible singularities at $u=v$.
All the $+3$ states have the form, up to the Gaussian factor
$\exp\left(-\bar{u}u-\bar{v}v\right)$,
$$P(z_i,\bar{z}_i)\left[ (z_1-z_2)(z_2-z_3)(z_3-z_1) \right]^{\nu}
=P(u,\bar{u},v,\bar{v})(u^3-v^3)^{\nu}$$
with $P$ a polynomial and $\nu\ge0$.
Therefore the only source of singularity can be
differentiation of the last factor. This immediately
implies that $a_u^{\dagger}$, which involves
${\partial}_{\bar u}$ only, is always regular, and
so is $a_v^{\dagger}$; hence, $c^{\dagger}_{22}$ and
$c^{\dagger}_{33}$ are regular. Now,
$$ b_u^{\dagger} (u^3-v^3)^{\nu} = \frac{1}{\sqrt{2}}\bar{u}
(u^3-v^3)^{\nu} - \frac{3\nu}{\sqrt{2}} u^2 (u^3-v^3)^{\nu-1}
$$
shows that $b_u^{\dagger}$ by itself may be singular, due to
the last term. Nevertheless,
\begin{eqnarray}
c^{\dagger}_{20} (u^3-v^3)^{\nu} &\!\!\! = \!\!\!& \mbox{regular
terms} -
\frac{3\nu}{2} u^3 (u^3-v^3)^{\nu-1}
+ \frac{3\nu}{2} v^3 (u^3-v^3)^{\nu-1}
\nonumber \\ &\!\!\! = \!\!\!& \mbox{regular terms}
- \frac{3\nu}{2} (u^3-v^3)^{\nu} \nonumber
\end{eqnarray}
is regular. Further,
$$
c^{\dagger}_{31} (u^3-v^3)^{\nu} = \mbox{regular terms} -
\frac{3\nu}{2} u^2 v^2 (u^3-v^3)^{\nu-1}
+ \frac{3\nu}{2} v^2 u^2 (u^3-v^3)^{\nu-1},
$$
but $c^{\dagger}_{3,-3}$, $c^{\dagger}_{2,-2}$
and $c^{\dagger}_{3,-1}$ are singular.
A generic (not normalized) linear state of slope $s=+3$
may then be written as
\begin{eqnarray}
\psi_{klmn}(u,v;\nu)
&\!\!\!=\!\!\!& (c_{20}^\dagger)^k (c_{31}^\dagger)^l
(c_{22}^\dagger)^m
(c_{33}^\dagger)^n \psi_{0}(u,v;\nu) \nonumber \\
&\!\!\!=\!\!\!& (c_{20}^\dagger)^k (c_{31}^\dagger)^l
(c_{22}^\dagger)^m
(c_{33}^\dagger)^n (u^3 - v^3)^\nu e^{-\bar{u}u-\bar{v}v}, \label{39}
\end{eqnarray}
with
\begin{eqnarray}
E &\!\!\!=\!\!\!& 2k+3l+2m+3n+3\nu+2, \label{40} \\
L &\!\!\!=\!\!\!& l+2m+3n+3\nu. \label{41}
\end{eqnarray}
(For the $-3$ states, of course, $c^{\dagger}_{p,-q}$
would take the place of $c^{\dagger}_{pq}$.)
This is a classification of {\it all\/} linear three-anyon
states (cf.~\cite{SporreNP,Dunne,book}).
However, some of these states are tower excitations and some
are different parts of the same trajectories.
First, the operator $c_{20}^\dagger$ is nothing but the tower raising
operator. It never produces singularities when acting on
any state, linear or nonlinear \cite{Sen92},
thus providing the tower structure of the whole spectrum.
Hence put $k=0$. Second, note that
$\psi_{00mn}(u,v;\nu) =
2^{m+3n/2} (uv)^m (u^3+v^3)^n (u^3-v^3)^\nu e^{-\bar{u}u-\bar{v}v}$
and consequently
\begin{eqnarray*}
\psi_{0,0,m,n+2}(u,v;\nu) - 4\psi_{0,0,m+3,n}(u,v;\nu) &\!\!\! =
\!\!\!&
8[(u^3+v^3)^2 - 4(uv)^3] \psi_{00mn}(u,v;\nu) \\
&\!\!\! = \!\!\!& 8(u^3-v^3)^2 \psi_{00mn}(u,v;\nu) \\
&\!\!\! = \!\!\!& 8\psi_{00mn}(u,v;\nu+2);
\end{eqnarray*}
in other words, the wave functions (\ref{39}), being linearly
independent for $0 \le \nu < 2$, are no longer so for
$0 \le \nu < \infty$, therefore some of them do not lead to
new trajectories and should be excluded from the trajectory
counting. As the last formula shows, it is sufficient to restrict to
$n < 2$.
Thus, a bottom trajectory may be labeled by three numbers
$l,m = 0,1,2,\ldots$ and $n=0,1$.
This is completely equivalent to the labeling by $(E_+,L_+)$,
because it follows from (\ref{40})--(\ref{41}) that
\begin{eqnarray}
E_+ &\!\!\!=\!\!\!& 3l+2m+3n+2, \label{45} \\
L_+ &\!\!\!=\!\!\!& l+2m+3n, \label{46}
\end{eqnarray}
since the point of slope change to $+3$ here is $\nu=0$,
and it is straightforward to see that for
any point $(E_+,L_+)$ there is no more than one set $\{lmn\}$ such
that these two equations are satisfied.

The set of bottom trajectories being {\it two\/}-parametric
(plus a ``double degeneracy'' due to $n$) is due to the fact that a
bottom state is identified by {\it three\/} quantum numbers (six
degrees
of freedom, minus two for the center of mass and one for the tower
excitations). Two quantum numbers (say, $E_+$ and $L_+$) identify
a trajectory, and the third one ($L$) chooses a point on that
trajectory. To compare, in the two-anyon problem, only {\it one\/}
quantum number, $L$, is enough to identify a bottom state,
consequently there is only one trajectory. In general, for the
$N$-anyon problem, the family of trajectories will be
$(2N-4)$-parametric.
Let us return once more to Fig.~1; there are two copies of one and
the same two-dimensional pattern, made of bullets ($n=0$) and
triangles ($n=1$), and increasing $m$ or $l$ by 1 means moving within
the pattern (3 units to the right and 3 units up, or 1 unit to
the right and 3 units up, respectively).

Thus the counting of trajectories is complete, but to find the
wave functions, certain modifications of eq.~(\ref{39}) are still
necessary. First, we need only such functions for which
$c_{20}\psi=0$,
where $c_{20} = a_ub_u + a_vb_v$ is the tower lowering operator,
but since $c_{20}$ does not commute with the $c^\dagger$'s,
even for $k=0$ some of the functions (\ref{39}) will contain
an admixture of non-bottom states with the same $E$ and $L$.
To correct for this, it turns out to be sufficient
to replace $(c_{31}^\dagger)^l$ in (\ref{39}) by another operator,
writing
\begin{eqnarray}
\psi_{lmn}(u,v;\nu) &\!\!\!=\!\!\!& C_l^\dagger (c_{22}^\dagger)^m
(c_{33}^\dagger)^n \psi_{0}(u,v;\nu) \nonumber \\
&\!\!\!\propto\!\!\!& C_l^\dagger (uv)^m (u^3+v^3)^n (u^3-v^3)^\nu
e^{-\bar{u}u-\bar{v}v}, \label{42}
\end{eqnarray}
where
\begin{equation}
C_l^\dagger = Q^l [(a_u^\dagger)^3-(a_v^\dagger)^3]^l
\label{43}
\end{equation}
and
\begin{equation}
Q = a_u b_v^\dagger - a_v b_u^\dagger
\label{44}
\end{equation}
is Sen's supersymmetry operator \cite{Sen,Sen92}.
Clearly, $C_l^\dagger$ is symmetric, and it is straightforward
to check that $c_{20}\psi_{lmn}=0$ always [$c_{20}$ commutes
with $Q$ and gives
zero when acting on $f(u,v)e^{-\bar{u}u-\bar{v}v}$].
Also, $C_l$ changes $E$ and $L$ by the same amount as
$(c_{31}^\dagger)^l$, so that the formulas (\ref{45}) and (\ref{46})
remain valid.

Second, a wave function that is continuous at the point
$(E_+,L_+)$---and
thus corresponds to a trajectory---is, in general, not one of the
form
(\ref{42}) but a linear combination of several functions of that
form.
Indeed, for any fixed $l,m,n$ and $0 \le \nu <2$ we have $(d+1)$
degenerate $s=+3$ states $\{ \psi_{l,m-3j,n} (u,v;\nu+2j), \:
j=0,1,\ldots,d \}$, where $d=[m/3]$. These are linearly independent,
but not orthogonal; and which of their linear combinations to choose
as a basis in the relevant $(d+1)$-dimensional subspace, makes no
difference anywhere except at the point $\nu=0$. Here,
the states with $j=1,\ldots,d$ continue as linear to $\nu<0$,
and the basis in
the relevant $d$-dimensional subspace may still be chosen at will,
but one state ``decouples'', i.e., becomes nonlinear,
and that is the one orthogonal to this subspace. In general, it
will be not $\psi_{lmn}(u,v,0)$ but a sum $\sum_{j=0}^d a_j
\psi_{l,m-3j,n} (u,v;\nu+2j)$, where the numbers $a_j/a_0$,
$j=1,\ldots,d$, are determined by the $d$ orthogonality conditions.
Consider the simplest example $l=0$, $m=3$, $n=0$, which corresponds
to the bosonic point $(8,6)$, see Fig.~2c. The two relevant states
of the form (\ref{42}) are $\psi_I \equiv \psi_{030}(u,v;\nu)$ and
$\psi_{II} \equiv \psi_{000}(u,v;\nu+2)$. At $\nu>0$, it makes no
difference which two linear combinations of them to choose;
but at $\nu=0$, $\psi_{II}$ continues to be linear, while the
state which actually ``decouples'' and continues to $\nu<0$ as
nonlinear,
is the one orthogonal to $\psi_{II}$ at $\nu=0$. An elementary
calculation shows that it is $\psi_I + \frac{4}{11} \psi_{II}$.
Such reasoning has to be repeated at each bosonic point with
$m \ge 3$, and general expressions for the coefficients $a_j$
apparently do not exist.

\section{The nonlinear parts of the trajectories}

We have shown that each trajectory is uniquely
identified by its $(E_+,L_+)$ point; the point being
given, one should in principle be able to reconstruct
the behavior of the whole trajectory, and in particular
to find all the bosonic and fermionic points it passes
through. In fact, it would be sufficient to find
$(E_-,L_-)$ and the behavior between that point and
$(E_+,L_+)$ [where $E(L)$ is nonlinear].
To see whether this can be achieved, we plot the
low-lying trajectories, using the exact multiplicities
given in \cite{Mash93}, on Fig.~2a--f. (There are six
different values of $L(0)$ and consequently six plots, one for
each class of trajectories.)
The nonlinear pieces are shown schematically only.
The continuity of the trajectories is confirmed by
the fact that at bosonic and fermionic points, where
they cross, the total number of trajectories coming from
the left and from the right is always the same. The
main question is how to identify which piece on the
left is a continuation of which piece on the right.
If at a certain crossing point perturbation theory
works and the first-order corrections to energy are
different for all the states involved, then the rule is
that the order of trajectories, by increasing energy, on
the right is opposite to that on the left (because the
first-order corrections have opposite signs). This argument
does not work (a) if the first-order corrections to some
of the states are equal and (b) when perturbation theory
breaks down, at certain bosonic points, as mentioned above.
A way out then is to find the wave functions numerically, using
the method of \cite{3ansol} and to identify them by comparing
their limits as the crossing point is approached. As it
appears, the above rule---the trajectory which is the $n$-th
from above on the left is the $n$-th from below on the right---holds
for all crossings analyzed. In other words, all crossings appear
to be {\it true\/} crossings.
When this rule is used, it becomes possible
to identify the trajectories completely without actually
finding the wave functions.
It turns out that for all the trajectories there is at most one
point of slope change, apart from
$(E_+,L_+)$ and $(E_-,L_-)$, which we denote by
$(E_0,L_0)$, the change being from $-1$ to $+1$
as $L$ increases through $L_0$; clearly,
$L_- \le L_0 \le L_+$. A generic behavior of a trajectory is,
therefore,
$$ \stackrel{-3}{\rightarrow} (E_-,L_-)
\stackrel{-1}{\rightarrow} (E_0,L_0)
\stackrel{+1}{\rightarrow} (E_+,L_+)
\stackrel{+3}{\rightarrow} $$
(numbers on top of the arrows meaning slopes), but in fact it
may be $L_0=L_-$ or $L_0=L_+$ (for the ground state,
$L_-=L_0=L_+=0$).

Table 1 shows the points of slope change for all
the trajectories with $E_+ \le 12$.
\begin{center}
\begin{tabular}{|rr|rr|rr|c|rr|rr|rr|}
\cline{1-6} \cline{8-13}
$E_- \!\!$ & $L_- \!\!$ & $E_0 \!\!$ & $L_0 \!\!$ & $E_+ \!\!$ & $L_+
\!\!$ &&
$E_- \!\!$ & $L_- \!\!$ & $E_0 \!\!$ & $L_0 \!\!$ & $E_+ \!\!$ & $L_+
\!\!$ \\
\cline{1-6} \cline{8-13}
 2 &0    & 2&0     & 2&0  & & 9 &$-$5 & 7 &1    & 9 &7  \\
 6 &$-$4 & 4&2     & 4&2  & & 10&$-$8 & 8 &$-$2 & 10&4  \\
 7 &$-$5 & 5&1     & 5&1  & & 10&$-$6 & 8 &0    & 10&6  \\
 5 &$-$3 & 4&0     & 5&3  & & 10&$-$4 & 8 &2    & 10&8  \\
 4 &$-$2 & 4&$-$2  & 6&4  & & 11&$-$3 & 10&0    & 11&3  \\
 11&$-$9 & 7&3     & 7&3  & & 13&$-$7 & 10&2    & 11&5  \\
 5 &$-$1 & 5&$-$1  & 7&5  & & 13&$-$11& 9 &1    & 11&7  \\
 8 &$-$4 & 7&$-$1  & 8&2  & & 7 &$-$3 & 7 &$-$3 & 11&9  \\
 8 &$-$2 & 7&1     & 8&4  & & 14&$-$12& 10&0    & 12&6  \\
 8 &$-$6 & 6&0     & 8&6  & & 14&$-$10& 10&2    & 12&8  \\
 9 &$-$7 & 7&$-$1  & 9&5  & & 14&$-$8 & 10&4    & 12&10 \\
\cline{1-6} \cline{8-13}
\end{tabular}
\end{center}
\nopagebreak
\begin{center}
Table 1. The behavior of the low-lying trajectories.
\end{center}
The regularity is not obvious here, but it becomes visible
when one looks at sufficiently high-lying trajectories.
Introduce the quantities
\begin{equation}
n_+ = \frac{L_+-L_0}{6}, \qquad n_- = \frac{L_0-L_-}{6},
\label{p31}
\end{equation}
that is, the ``numbers of revolutions'', or the numbers of bosonic
points that a trajectory passes through having slopes $+1$ and
$-1$, respectively; these may be integer or half-integer, as $L_0$
may correspond to either a bosonic or a fermionic point. The
following obvious relations hold
\begin{eqnarray}
E_0 &\!\!\!=\!\!\!& E_+ - 2n_+, \phantom{({}-n_-)} \quad
L_0 = L_+ - 6n_-, \label{p32} \\
E_- &\!\!\!=\!\!\!& E_+ - 2(n_+-n_-), \quad L_- = L_+ - 6(n_+ + n_-).
\label{p33}
\end{eqnarray}
The results, deduced from the numerical analysis, are the following:
\begin{eqnarray}
\hspace*{-2em}\frac{E_+}{3} < L_+ < \frac{E_+}{2}: &\;&
 n_+ = \frac{3L_+ - E_+}{4}, \quad n_- = \frac{E_+ - 2L_+}{2},
 \quad L_0 \mbox{ is fermionic,} \label{n+} \\
\hspace*{-2em}\frac{E_+}{2} < L_+ < E_+: &\;&
 n_+ = \frac{E_+}{8},\phantom{3L_+ -{}} \quad  n_- =
\frac{E_+}{8},\phantom{{}-2L_+}
 \quad L_0 \mbox{ is bosonic.} \label{n-}
\end{eqnarray}
The formulas, as well as the inequality signs, are valid
asymptotically
for $E \gg 1$, in the same sense as (\ref{b+3})--(\ref{b+1}).
One can verify the compatibility of (\ref{n+})--(\ref{n-}) with
(\ref{b+3})--(\ref{b+1});
also, a simple consistency check comes from the symmetry
requirement---a
trajectory $(E_-,L_-) \to (E_0,L_0) \to (E_+,L_+)$ must have its
partner $(E_+,-L_+) \to (E_0,-L_0) \to (E_-,-L_-)$: this is
satisfied as well. Note that $n_+$ depends continuously on $L_+$, but
there is a discontinuity in $n_-$ at the point $L_+ = E_+/2$.
Also, the exact results exhibit certain periodic structure
with period 8 by $E_+$ and 4 by $L_+$. Explaining these features
remains an open item.

A semiclassical interpretation---rather rough, although---of some
features of the trajectories may be adduced. When two particles
are close together and the third one is far from them, that is,
say, $|\rho_{12}| \ll |\rho_3|$ [$\rho_{12} = z_1 - z_2, \;
\rho_3 = (z_1+z_2)/2 - z_3$], one gets two independent
anyonic oscillators, one ($\rho_{12}$) with the statistics parameter
$\nu$, the other ($\rho_3$) with the parameter $2\nu$
\cite{SporreNP,Syst}. The energy in this approximation is
$|2l_{12} + \nu| + 2n_{12} + 1 + |2l_3 + \nu| + 2n_3 + 1$,
where in general $l_3 \gg l_{12}$; as $\nu$ increases through
$-2l_{12}$, the slope changes from $-3$ to $-1$ (if $l_3<0$)
or from $+1$ to $+3$ (if $l_3>0$); the other two points of
change, however, can not be described correctly, nor
can the formulas (\ref{n+})--(\ref{n-}) be explained.
Perhaps a more accurate
semiclassical approximation would be able to explain them.

\section{Conclusion}

We have developed the concept of trajectories
for anyons, noting that many different many-anyon
states are in fact continuations of each other.
We discuss in particular the harmonic oscillator
external potential.
Apart from the center-of-mass and tower excitations,
in the two-anyon problem there is only one trajectory.
For the three-anyon problem, we have worked out the classification
and the main features of the trajectories: The slope of a
trajectory always changes as $-3 \to -1 \to +1 \to +3$
(where the $+1$ and/or $-1$ pieces may be missing), and
the point $(E_+,L_+)$ of the last change may be used to label
the trajectory. The wave function corresponding to the
linear part of the trajectory can be written down exactly by
applying excitation operators to the ground trajectory.
Concerning the nonlinear part, we conjecture the formulas,
based on a numerical analysis, expressing the lengths of
intervals of slope $-1$ and $+1$ in terms of $(E_+,L_+)$.

Qualitatively, the picture will be the same for the
$N$-anyon problem: Again, each trajectory will have
the points $(E_+,L_+)$ and $(E_-,L_-)$ such that
its $E(L)$ dependence is linear with slopes
$\pm N(N-1)/2$ for $L>L_+$ ($L<L_-$) and
nonlinear in between. Here the family of trajectories
will be $(2N-4)$-parametric.
It is quite plausible that in this case as well, the sequence
of slopes will be regular, from $-N(N-1)/2$
up to $+N(N-1)/2$ in steps of two. However,
to gain a more precise understanding of the behavior
of the trajectories remains an open problem.

\bigskip

S.~M. would like to express his sincere thanks to the
Institutt for fysikk at the NTH, Trondheim,
and the theory division
of the Institut de Physique Nucl\'eaire, Orsay,
where parts of this work were done,
for the kind hospitality and support.

\newpage
{\bf Figure captions}
\bigskip

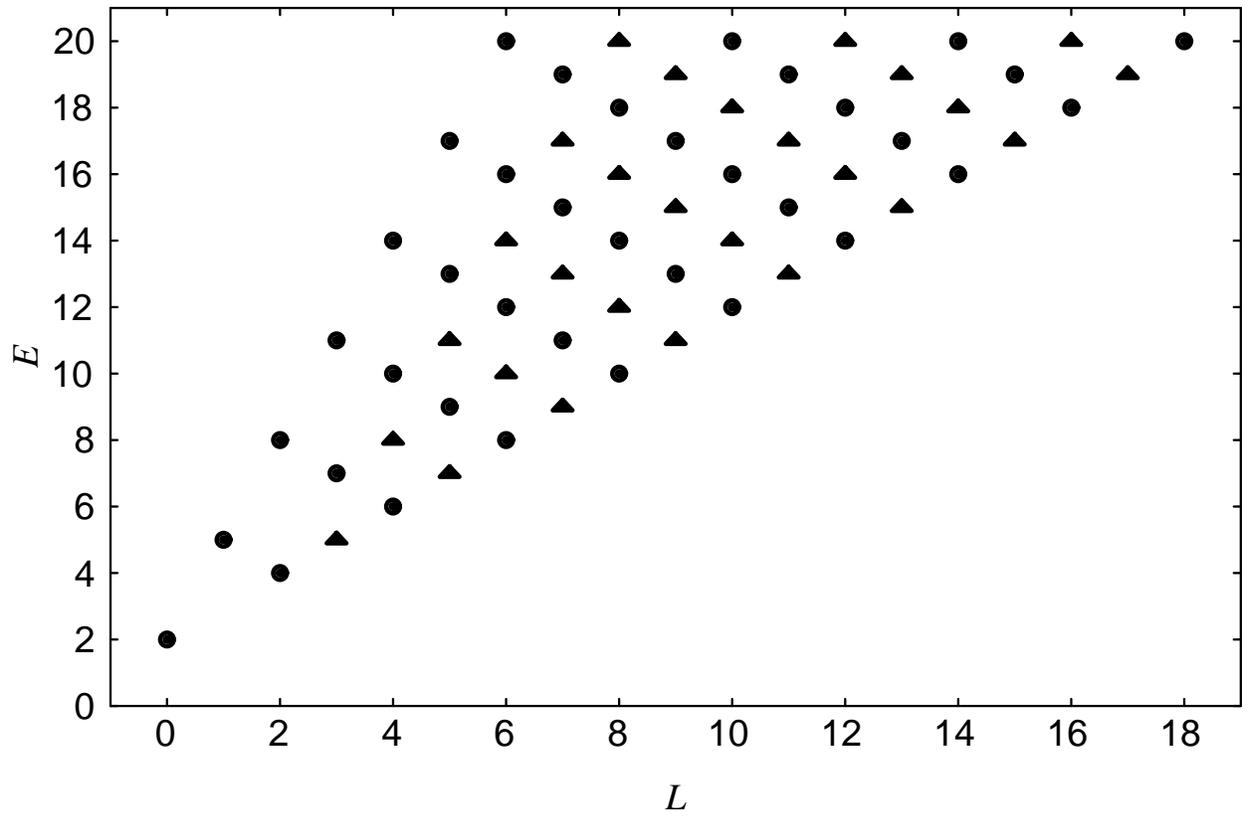
Fig.~1. The distribution in the $(E,L)$ plane of the points
        $(E_+,L_+)$ at which the trajectories turn linear with
        slope $s=+3$. The bullets define the set
        $E_+=3\ell+2m+2$, $L_+=\ell+2m$, and the triangles define
        the set $E_+=3\ell+2m+5$, $L_+=\ell+2m+3$, with
        $\ell=0,1,\ldots$ and $m=0,1,\ldots$ in both cases.

Fig.~2, a--f. The six classes of trajectories in the $(E,L)$ plane,
        with $L(0)=-2,-1,0,1,2,3$; solid and
        dashed vertical lines mean Bose and Fermi statistics,
        respectively, dashes show the multiplicities of the linear
        states.
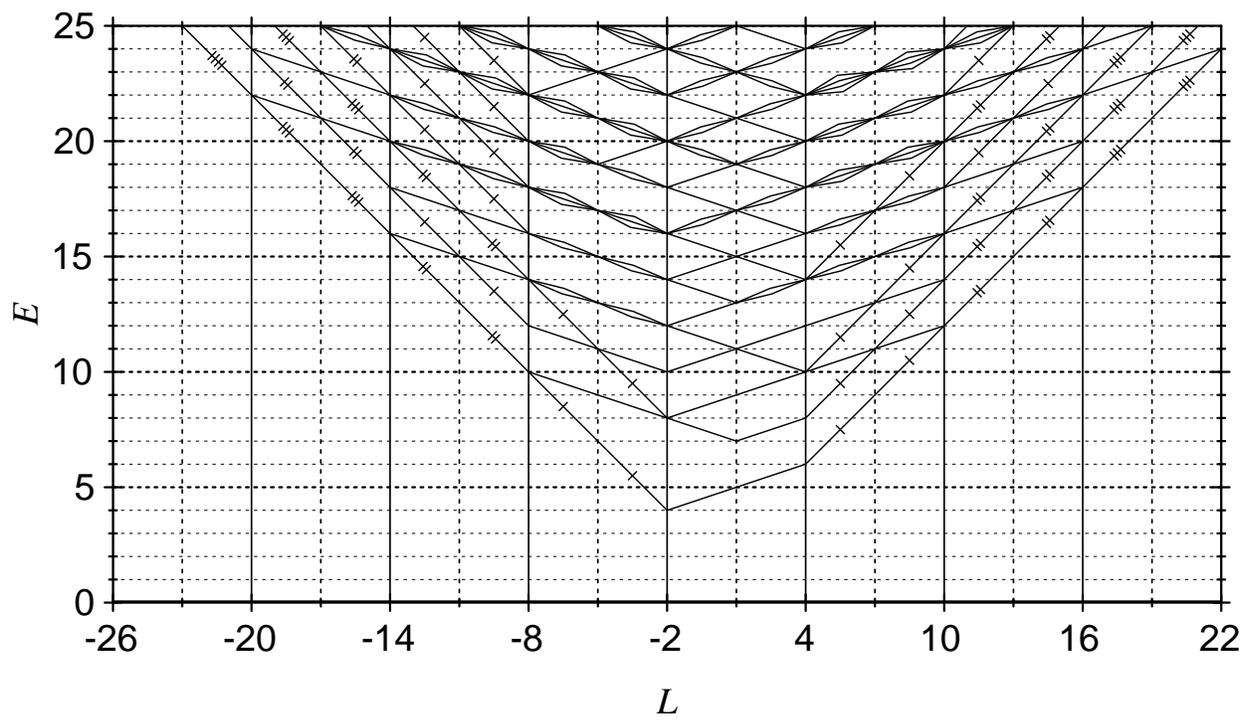
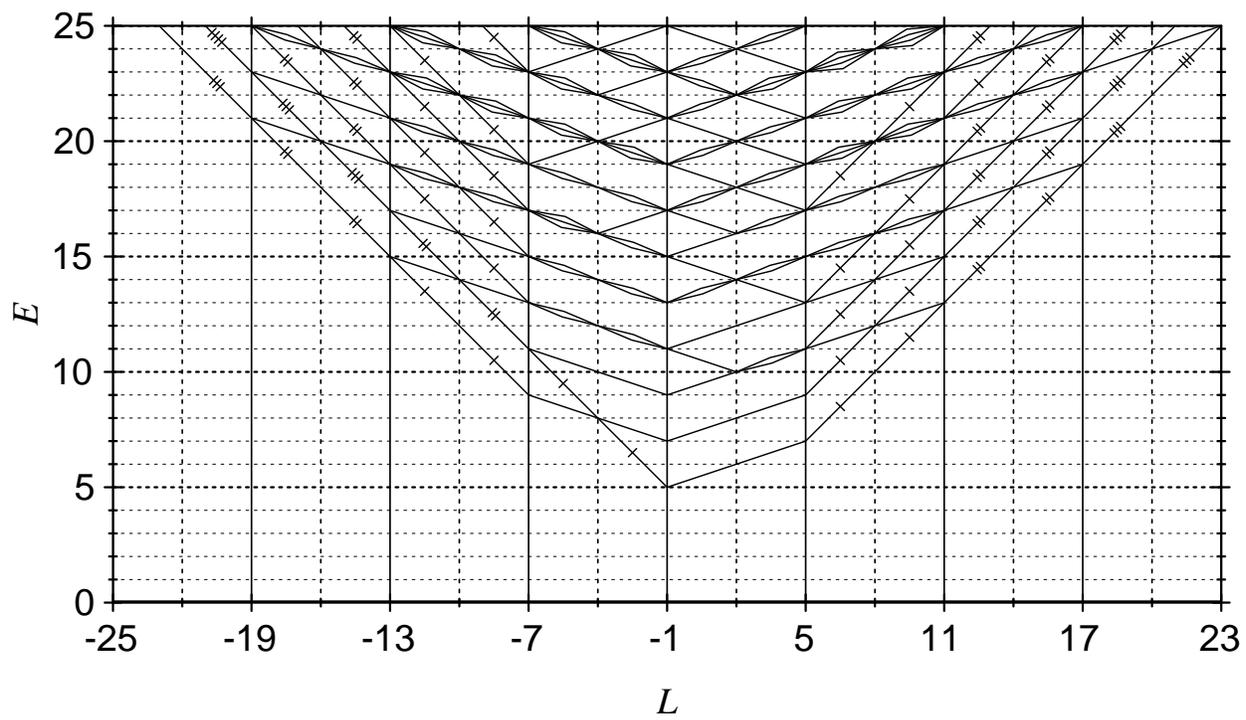
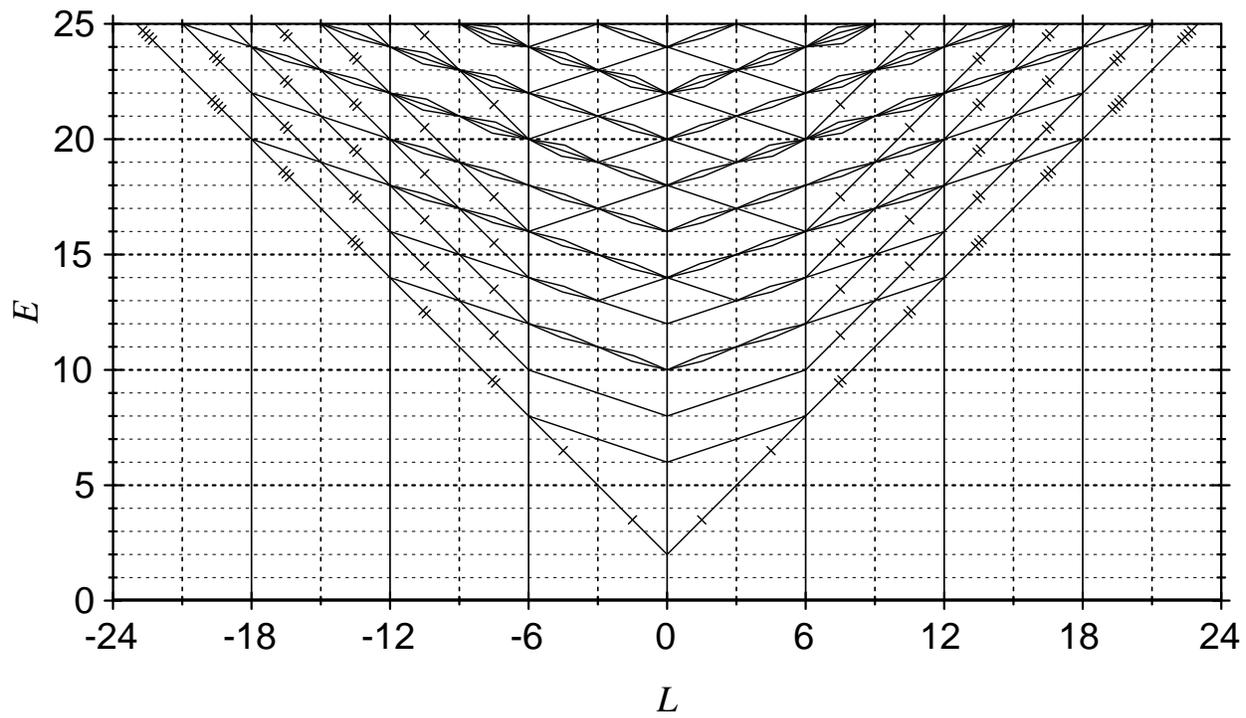
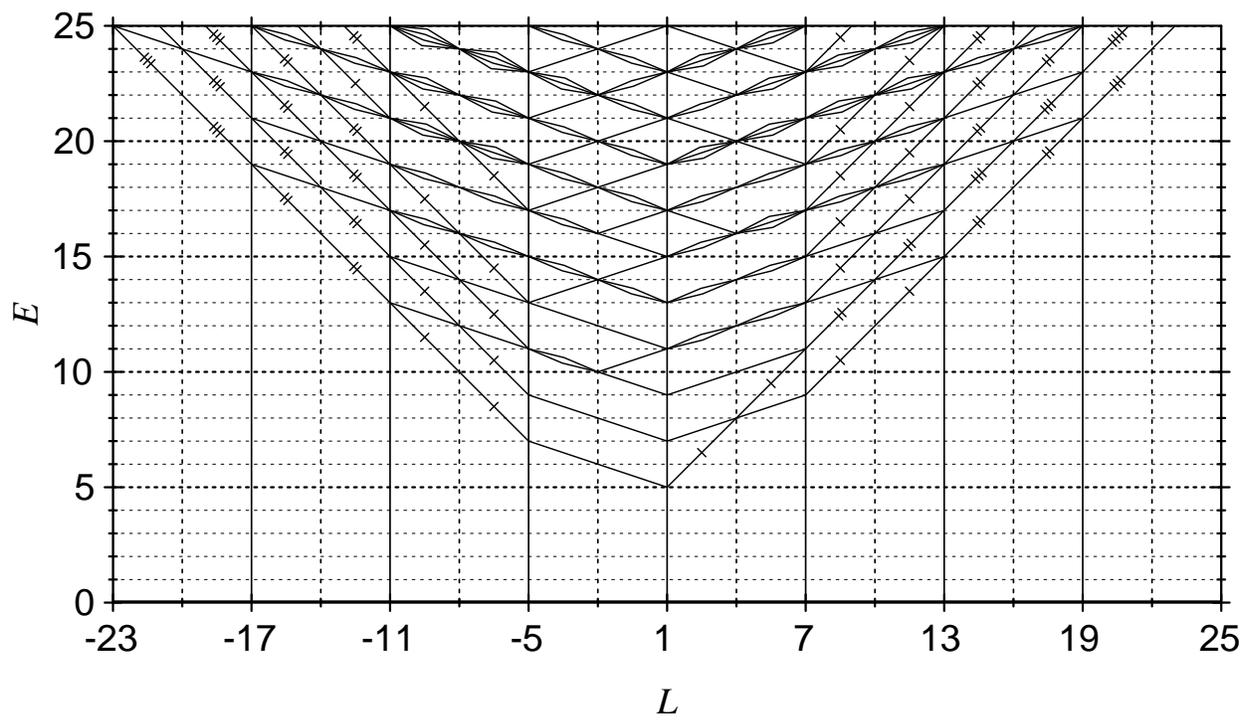
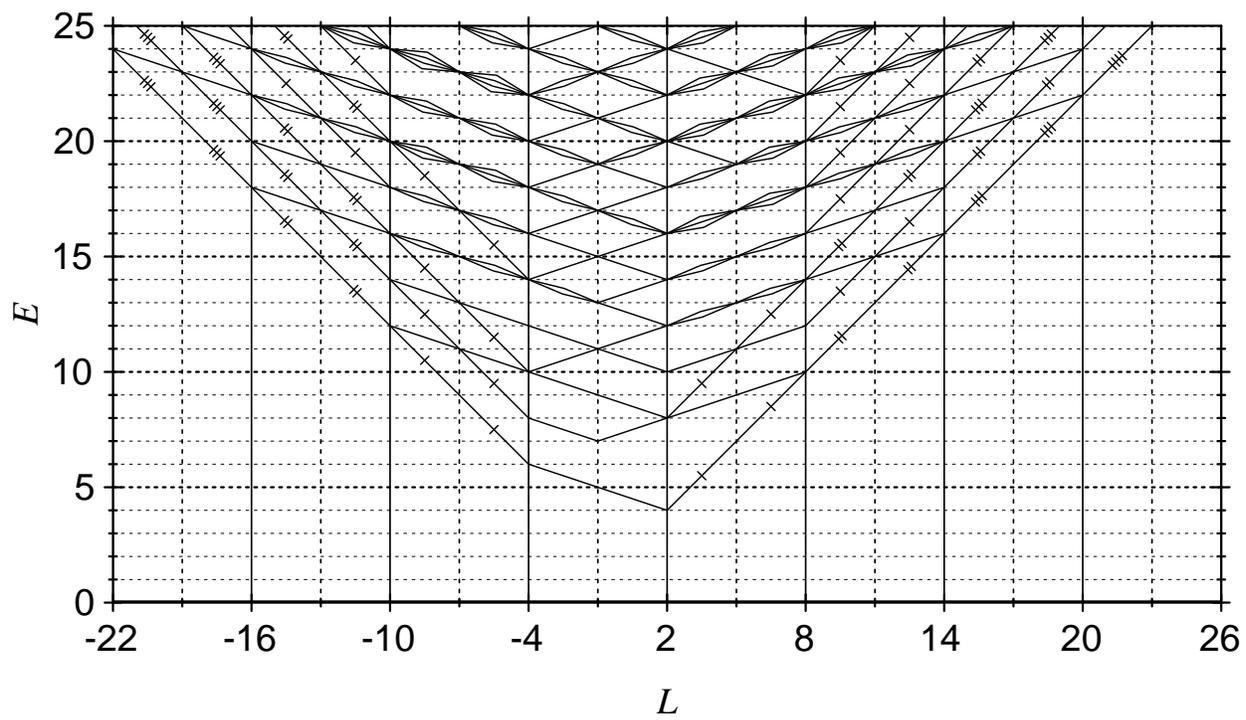
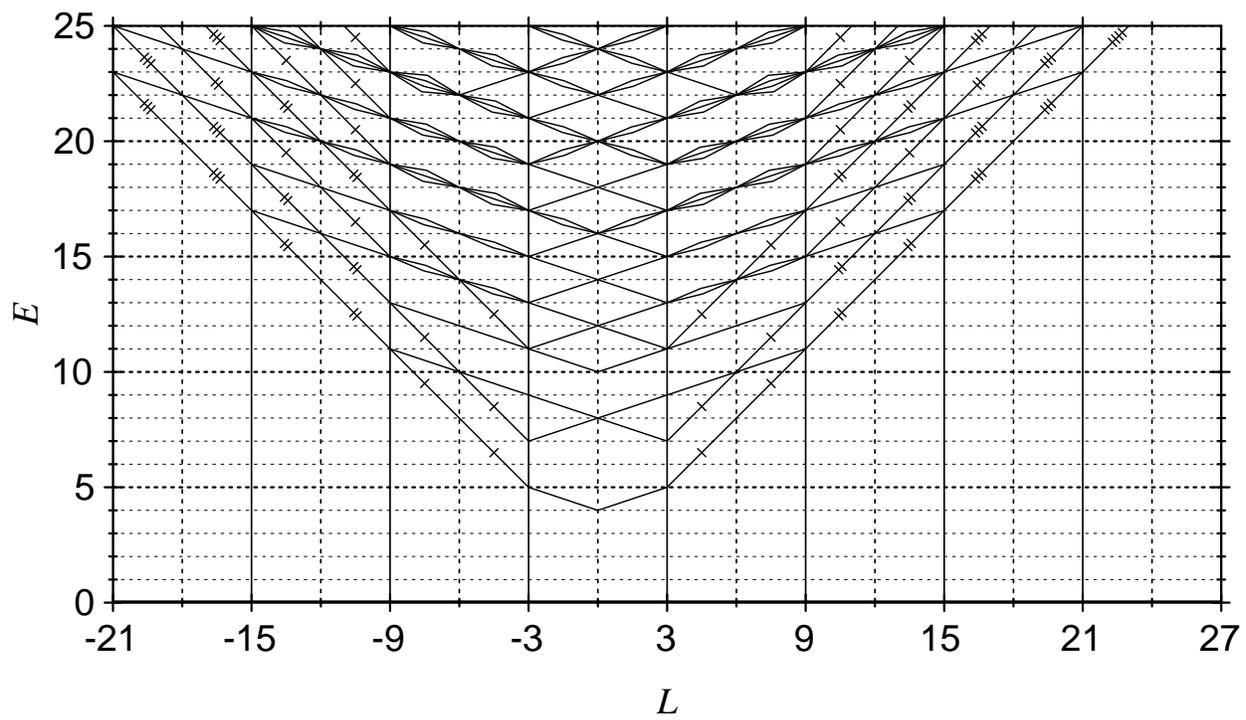
\end{document}